# Investigating Thermal Cooling Mechanisms of Human Body Under Exposure to Electromagnetic Radiation

Huan Huan Zhang[1], Member, IEEE, Ying Liu[1], Senior Member, IEEE, Xiaoyan Y. Z. Xiong[2], Member, IEEE, Guang Ming Shi[3], Senior Member, IEEE, Chun Yang Wang[4], and Wei E.I. Sha[5], Senior Member, IEEE

[1]National Key Laboratory of Antennas and Microwave Technology, School of Electronic Engineering, Xidian University, Xi'an, 710071, China.
[2]ZJUI Institute, International Campus, Zhejiang University, Haining, 314400, China.
[3]School of Artificial Intelligence, Xidian University, Xi'an, 710071, China.
[4]College of Clinical Medicine, Xi'an Medical University, Xi'an, 710021, China.
[5]College of Information Science and Electronic Engineering, Zhejiang University, Hangzhou, 310058, China.

Corresponding author: Ying Liu (e-mail: liuying@mail.xidian.edu.cn).

The authors would like to thank the support of National Natural Science Foundation of China under Grant 61701376, 81702010, China Postdoctoral Science Foundation under Grant 2017M613071, 2018T111016, Open Project Foundation of the State Key Laboratory of Millimeter Waves under Grant K201808, Open Project Foundation of Key Laboratory of High-Speed Circuit Design and EMC, Ministry of Education.

**ABSTRACT** Thermal cooling mechanisms of human exposed to electromagnetic (EM) radiation are studied in detail. The electromagnetic and thermal co-simulation method is utilized to calculate the electromagnetic and temperature distributions. Moreover, Pennes' bioheat equation is solved to understand different thermal cooling mechanisms including blood flow, convective cooling and radiative cooling separately or jointly. Numerical results demonstrate the characteristics and functions for each cooling mechanism. Different from the traditional view that the cooling effect of blood is usually reflected by its influence on sweat secretion and evaporation, our study indicates that the blood flow itself is an important factor of thermal cooling especially for high-intensity EM radiation. This work contributes to fundamental understanding of thermal cooling mechanisms of human.

## I. INTRODUCTION

Humans and other endotherms have thermoregulation mechanisms to maintain their body temperature at a stable level even when their ambient temperature changes. The temperature of human body is physiologically divided into shell temperature and core temperature. Human beings have a core temperature of around 37℃, at which the human body systems operate in optimum [1]-[5]. With the extensive use of electromagnetic (EM) field and microwave technology, EM radiation becomes pervasive in people's daily life. The EM radiation can originate from household electrical appliance, mobile phones, mobile communication base station, medical equipment, high voltage electric power equipment, radar, laser, etc. The most immediate hazard of EM radiation is the increment of temperature of tissues and organs. If the body temperature deviates too much from the core temperature, it will have an adverse impact on organism. For instance, it is reported that high temperature will increase the blood-brain-barrier permeability, making it easier for the neurotoxic metabolites produced by peripheral organs and tissues to enter the brain. Thus the brain cells will be damaged when the temperature is above 40℃ [6]. When the temperature is above 42℃, a series of cytotoxic events will occur, such as protein denaturation, aberrant oxidative phosphorylation and impaired DNA synthesis, resulting in multiorgan failure and neuron non-reversible damage [7]. In severe cases, people may exhibit symptoms of hypotension, shock, heart failure, hydrocephalus and even death [8]-[10]. Therefore a thorough study of the thermal cooling mechanisms of human under exposure to EM radiation is both necessary and significant, which will establish a theoretical foundation for the prevention and therapy of thermal damage of human body.

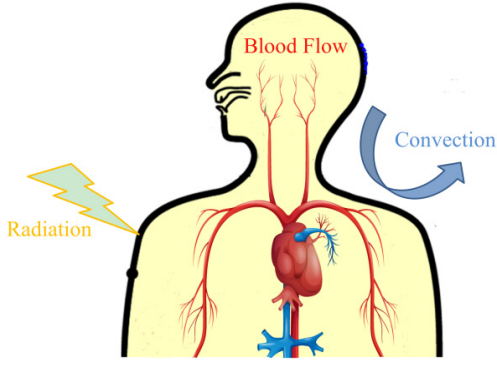

FIGURE 1. Cooling mechanisms of human body

In existing textbooks of medical physiology, the thermal cooling mechanisms of human body mainly include conduction, convection, radiation and evaporation [1], [2]. The heat in the internal of human body is transferred to the skin by thermal conduction. The skin can also conduct heat to whatever it is in direct contact with. Convection cooling occurs when the skin is surrounded by cool air. The air nearby the skin is warmed by the skin and its density decreases. As a result, the warmed air goes upward. Then the cool air in the vicinity will fill the vacancy and be warmed again. In such a manner, the heat of human body is transferred to the air. All objects that have a temperature higher than absolute zero radiate electromagnetic waves to surroundings. The temperature of human body is obviously higher than absolute zero. To transfer heat, human emits electromagnetic waves to surroundings through its skin. When the skin of human sweats out moisture, the evaporation of the liquid promotes heat loss [11].

Experimental study of the thermal cooling mechanisms on a living person is inconvenient for two reasons. It is difficult to conduct single factor analysis because the cooling mechanisms of human usually work in concert. Moreover, it is unethical to let a person stay under strong EM radiation or high ambient temperature. Consequently, in this work, a three-step electromagnetic and thermal co-simulation method [12]-[21] is adopted to study the thermal cooling mechanisms of human under exposure to EM radiation. Firstly, the finite element-iterative integral equation evaluation (FE-IIEE) method is utilized to solve the Helmholtz vector equation for the analysis of the electric field distribution inside a human head model beside an antenna [22], [23]. Secondly, the rate of electromagnetic energy absorbed by the biological tissues is represented by the specific absorption rate (SAR) value. At last, thermal simulation is performed to obtain the temperature distribution by solving the Pennes' bioheat equation with the time-domain finite element method (TDFEM) [24], [25]. The thermal cooling mechanisms of human is discussed in detail. A thorough study of the contributions of blood flow, convection and radiation for thermal cooling is carried out as shown in Fig. 1. Different from the traditional view that the cooling effect of blood is usually reflected by its influence on sweat secretion and evaporation, our study indicates that the blood flow itself is an important factor of thermal cooling.

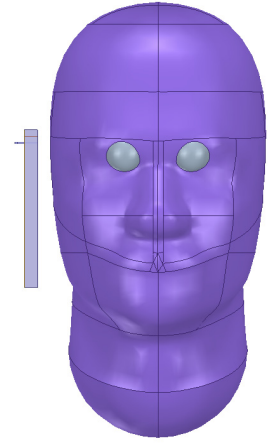

FIGURE 2. A human head model beside an antenna

TABLE I
ELECTROMAGNETIC AND THERMAL PARAMETERS OF HEAD MODEL

| Tissues | $\varepsilon_r$ | $\sigma$ | $\kappa$ | $c_\rho$ | $\rho$ | $Q_m$ | $B$ |
|---|---|---|---|---|---|---|---|
| Skin | 41.4 | 0.87 | 0.42 | 3500 | 1010 | 1000 | 9100 |
| Fat | 11.3 | 0.11 | 0.25 | 2500 | 920 | 180 | 520 |
| Muscle | 55.0 | 0.94 | 0.5 | 3600 | 1040 | 690 | 2700 |
| Bone | 12.5 | 0.14 | 0.4 | 1300 | 1810 | 0 | 1000 |
| CSF | 74.0 | 2.12 | 0.6 | 4000 | 1010 | 0 | 0 |
| Brain | 45.8 | 0.77 | 0.5 | 3630 | 1040 | 10000 | 35000 |
| Eye | 49.6 | 0.994 | 0.53 | 3615 | 1052 | 0 | 0 |

## II. METHODS

Firstly, the Helmholtz vector equation is solved to obtain the electric field distribution inside a human head model [22]:

$$\nabla \times \left( \mu_r^{-1} \nabla \times \mathbf{E} \right) - k_0^2 \varepsilon_r \mathbf{E} = 0, \qquad (1)$$

where $\mathbf{E}$ represents the electric field. $\mu_r$ refers to the relative permeability. $\varepsilon_r$ denotes the relative permittivity. $k_0$ is the wave number in free space.

Secondly, the SAR values are calculated to represent the rate of electromagnetic energy absorbed by the biological tissues:

$$SAR = \frac{\sigma \left| \mathbf{E}(\mathbf{r}) \right|^2}{\rho}, \qquad (2)$$

where $\sigma$ is the conductivity, and $\rho$ represents the density.

Thirdly, in order to understand the thermal cooling mechanisms of human, Pennes' bioheat equation is adopted for the thermal simulation [26]:

$$\rho c_\rho \frac{\partial T}{\partial t} = \nabla \cdot (\kappa \nabla T) + \rho SAR + Q_m - B(T - T_{blood}), \qquad (3)$$

where $T$ is the unknown temperature of a tissue as a function of time and space. $c_\rho$ refers to the specific heat capacity. $\kappa$ is the thermal conductivity. $Q_m$ denotes the metabolic heat generation. $T_{blood}$ is the blood temperature, and it is always close to the core temperature of human. $B$ is related to the blood perfusion.

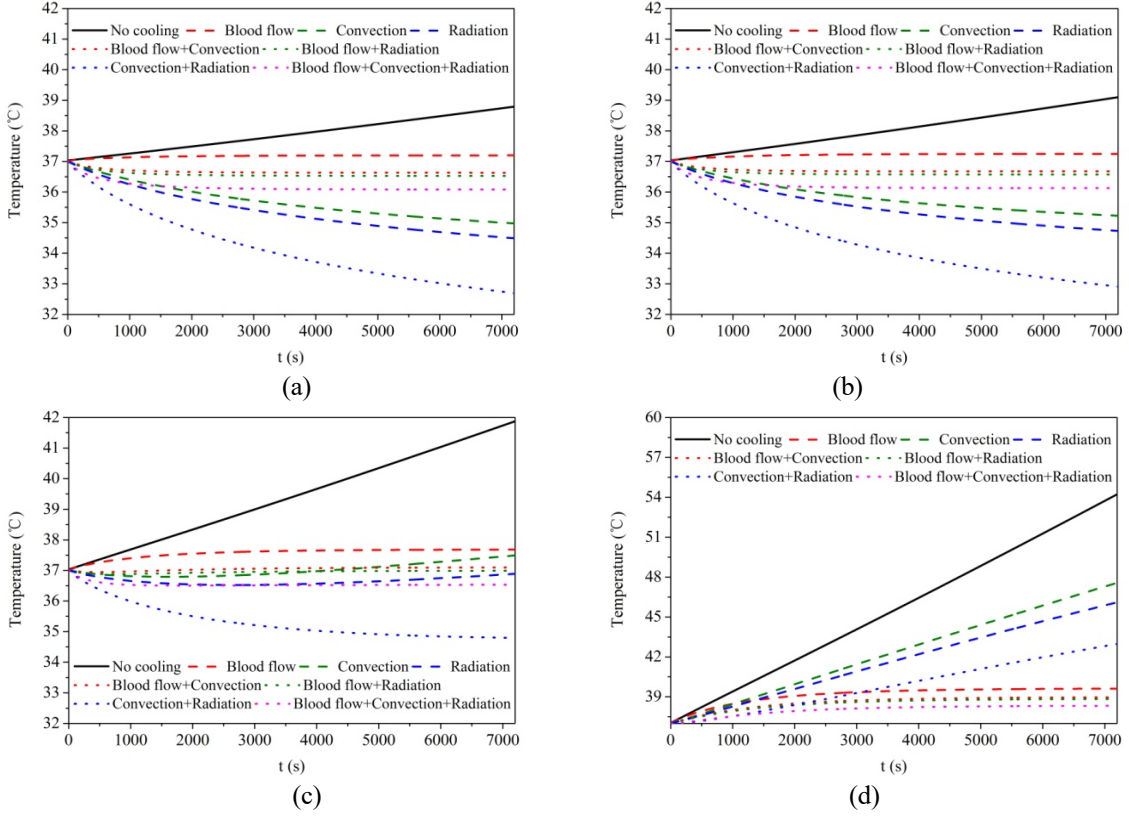

FIGURE 3. Real-time average body temperatures at ambient temperature=25℃, the input power of the antenna equals (a) 0 W (b) 1 W (c) 10 W (d) 50 W

A detailed description of the above three-step electromagnetic and thermal co-simulation method can be found in [12]. Actually, equation (3) is a revised version of traditional heat conduction equation by replacing the heat source with metabolic heat source and electromagnetic heat source [27], [28]. Meanwhile, the last term $-B(T-T_{blood})$ related to the blood flow is added to represent the heat exchange between the human tissues and blood. It can be observed that when the temperature of a tissue is higher than that of the blood, the last term is negative and the blood serves as a heat sink. The heat will be transferred from tissues to blood. Otherwise, the blood is a heat source. The heat will be transferred from blood to tissues. Thus the blood flow has very special thermal effects on the local tissue temperature.

Mathematically the differential equation (3) has unique solution only if its boundary conditions are given. The boundary of human body is the skin. Physically, heat transfer occurs between the skin and its surroundings through convection, radiation and evaporation. Due to the limitation of the Pennes' bioheat equation, evaporation is not considered in this work. The convection boundary condition and radiation boundary condition are adopted on the skin-air interface to describe the convection and radiation cooling mechanisms.

The convection boundary condition is expressed as
$$-\hat{n}\cdot\kappa\nabla T = h(T-T_{amb}), \qquad (4)$$
where $h$ represents the convective heat transfer coefficient, $T_{amb}$ denotes the ambient temperature. $\hat{n}$ denotes the unit outward normal vector of the boundary surface.

The radiation boundary condition is given by
$$-\hat{n}\cdot\kappa\nabla T = \varepsilon_0\sigma(T^4-T_{amb}^4), \qquad (5)$$
where $\varepsilon_0$ refers to the emissivity of the skin, $\sigma = 5.68\times 10^{-8}\ W/m^2 K^4$ is the Stefan-Boltzmann constant.

Solving the Pennes' bioheat equation with the specific boundary conditions, 1) we can study the thermal cooling of human body with or without external EM heat source at different ambient temperature. 2) we can capture the heat conduction from the internal of human body to the skin. 3) We can investigate the contributions of different cooling mechanisms involving convection, radiation and blood flow separately or jointly.

### III. NUMERICAL EXPERIMENTS and DISCUSSIONS
A human head model is exposed to the EM radiation of an antenna as shown in Fig. 2. The head model consists of skin, fat, muscle, cortical bones, cerebrospinal fluid (CSF), brain and eyes. The corresponding electromagnetic and thermal parameters of the tissues are given in Table I. The antenna is put beside the head model with a gap distance of 1 cm. Its working frequency is 900 MHz.

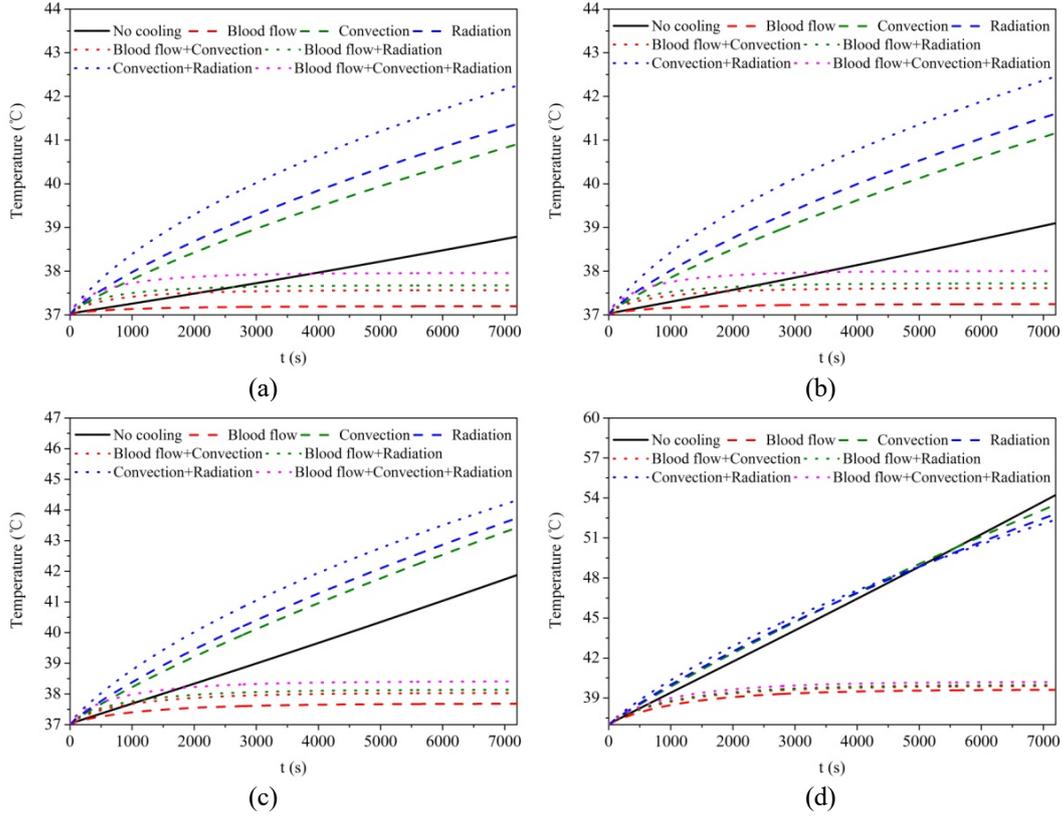

**FIGURE 4.** Real-time average body temperatures at ambient temperature=45℃, the input power of the antenna equals (a) 0 W (b) 1 W (c) 10 W (d) 50 W

The following assumptions are made before the cooling mechanisms of human are investigated:

1) The experimental subject is a normotensive male, who is awake at complete rest with no muscle activity and in a non-stressed state. Under such a condition, the metabolic rate $Q_m$ and the blood flow related term $B$ are given in Table I [13].
2) The EM radiation to the experimental subject is local. Thus the blood temperature can be the same as the core temperature due to blood circulation for a limited period.
3) The electromagnetic and thermal properties of human tissues are constant in our experiment.
4) The ambient temperature is homogeneous.
5) Free convection occurs between the skin and the air if convection cooling exists, $h = 5 W/(m^2 \cdot K)$ [29].
6) The emissivity of the skin of the experimental subject is $\varepsilon_0 = 0.95$ [29].

In the numerical experiment, $T_{amb}$=25℃ and $T_{amb}$=45℃ are considered corresponding to two situations that the ambient temperature is lower or higher than the body temperature of human. The input power of the antenna is set to be 0 W, 1 W, 10 W and 50 W to model different EM radiation sources. At a fixed ambient temperature and input power of the antenna, the average body temperatures are calculated by considering different thermal cooling mechanisms including convection, radiation, blood flow, and their combinations. The obtained real-time average body temperatures are shown in Fig. 3 and Fig. 4.

In Fig. 3 and 4, "No cooling" means that the skin surface is adiabatic. Convection cooling and radiation cooling do not exist. Meanwhile, there is also no blood flow in this case. Moreover, "Blood flow", "Convection" and "Radiation" suggest that there exists only one cooling mechanism. Furthermore, "Blood flow+Convection", "Blood flow+Radiation" and "Convection+Radiation" refer to that two of the three cooling mechanisms co-work. Additionally, "Blood flow+Convection+Radiation" corresponds to that all of the three cooling mechanisms take effect.

### A. No Cooling

Look at the black line in Fig. 3 and 4, the body temperature keeps rising due to the metabolic heat source and the EM radiation of antenna. Since the skin is adiabatic with the environment, the ambient temperature has no influence on the body temperature. So the results corresponding to "no cooling" in Fig. 3 and Fig. 4 are the same. At a fixed ambient temperature, the temperature rises faster when the input power of the antenna is larger.

### B. Only Convection or Radiation

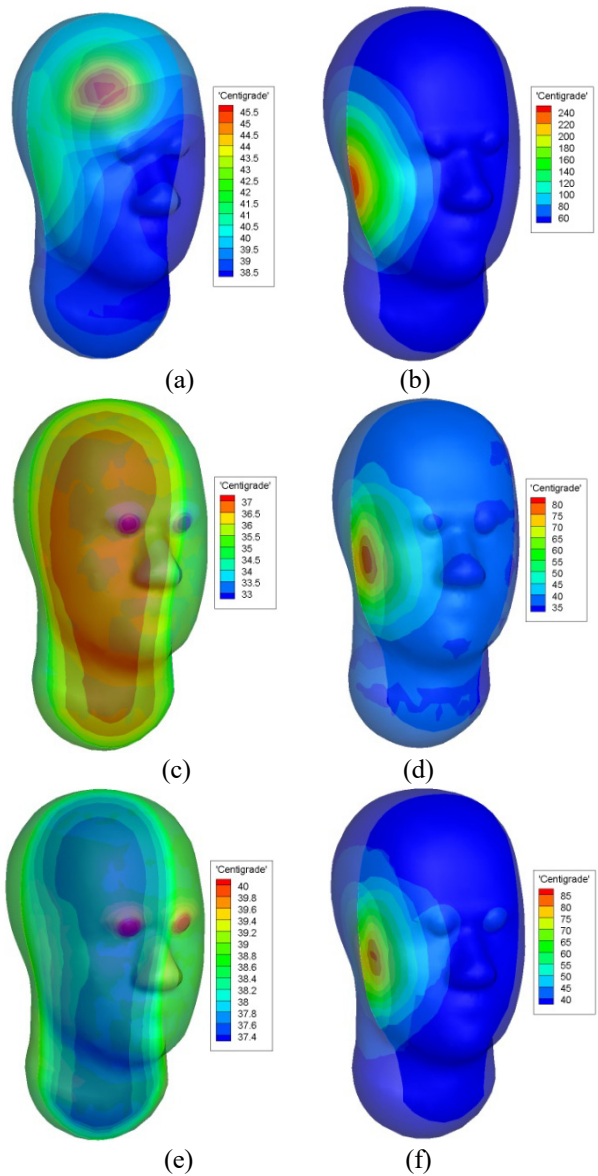

**FIGURE 5.** 3-D temperature distributions of human head model (a) Input power of 1 W, no cooling; (b) Input power of 50 W, no cooling; (c) Input power of 1 W, ambient temperature of 25℃, blood flow-convection-radiation combination; (d) Input power 50 W, ambient temperature of 25℃, blood flow-convection-radiation combination; (e) Input power of 1 W, ambient temperature of 45℃, blood flow-convection-radiation combination; (f) Input power of 50 W, ambient temperature of 45℃, blood flow-convection-radiation combination.

According to Fig. 3 (a)-(d), when only convection or radiation is considered, the body temperature decreases compared with the case of "no cooling" at ambient temperature of 25℃. Moreover, radiative cooling is more effective than convective cooling. According to Fig. 4 (a)-(c), both the convective cooling and radiative cooling lose efficacy at ambient temperature of 45℃ because the ambient temperature is higher than the body temperature. Heat will be transferred from the surroundings to human body through convection and radiation. The result of Fig. 4 (d) is interesting. There is an intersection point between the black line and the blue dashed line or between the black line and the green dashed line. Before the intersection point, the body temperature increases compared to the case of "no cooling" since the environment temperature is higher than the body temperature. After the intersection point, the body temperature is higher than the environment temperature. As a result, convective cooling and radiative cooling take effect. The body temperature becomes lower than that of "no cooling".

### C. Only Blood Flow

Use the case of "no cooling" as a baseline, when the body temperature is slightly higher than the core temperature as shown in Fig.3 (a)-(c), the cooling effect of blood flow is weaker than convection and radiation. However, when the input power of the antenna is large as shown in Fig. 3 (d), the body temperature is much higher than the core temperature. The cooling effect of blood flow becomes more remarkable than either of convection and radiation. This phenomenon is especially apparent when convective cooling and radiative cooling lose efficacy at ambient temperature of 45℃ in Fig. 4. Moreover, it can be observed that the blood flow does not continuously decrease the body temperature. It can make the body temperature stable around the normal core temperature of 37℃.

### D. Combination of Cooling Mechanisms

The case of "Convection+radiation" is similar with the case of only convection or radiation. Moreover, "Convection+radiation" decreases or increases the body temperature faster than only convection or radiation at the ambient temperature of 25℃ or 45℃.

The cases of "Blood flow+Convection", "Blood flow+Radiation" and "Blood flow+Convection+Radiation" are very similar. Take the case of "Blood flow+Convection" as an example and compare it to the case of "Convection". As shown in Fig.3 (a) and Fig.3 (b), "Convection" decreases the body temperature below 37℃. While with the consideration of blood flow, "Blood flow+Convection" can increase the body temperature above 37℃. In Fig. 3(c), there is an intersection point between the green dashed line and the red dotted line. Before the intersection point, the body temperature is decreased below 37℃ with "Convection", while it will be increased around 37℃ with "Blood flow+Convection". After the intersection point, "Convection" cannot offset the temperature increase induced by EM radiation and metabolic heat. The body temperature becomes higher than the core temperature. The blood flow starts to decrease the body temperature. So the body temperature of "Blood flow+Convection" is lower than "Convection" after the intersection point. In Fig.3 (d) and Fig.4 (a)-(d), "Convection" cannot effectively decrease the body temperature or will increase it. But after adding the contribution of blood flow, the cooling effect is very evident. Especially when the ambient temperature is 45℃ and the input power of the antenna is high, "Blood flow+Convection" can guarantee the body temperature not far from the core temperature. Based on the above observation, the blood will heat the body when the body temperature is lower than the core temperature, while it will cool the body when the body temperature is higher than the core temperature.

Figure 5 demonstrates the 3D temperature distributions at t=7200 s corresponding to "no cooling" or "Blood flow+Convection+Radiation" under the input power of antenna 1 W and 50 W at ambient temperature 25℃ and 45℃. It can be observed that the body temperature will be increased drastically due to high-power EM radiation if there is no cooling. If the power of the EM radiation is low, the temperature distribution is relatively uniform and the hotspot will be in the center of the head model. If the power of the EM radiation is high, the hotspot will approach the source of radiation. With the consideration of the all-combined cooling mechanisms, the body temperature can be decreased effectively. On one hand, under high EM radiation (50 W), the temperature of hotspots close to the EM source is still much higher than the core temperature. On the other hand, under low EM radiation (1 W), the Human's eyes are the most dangerous places above the core temperature.

## IV. CONCLUSION

In this paper, electromagnetic and thermal co-simulation is carried out to analyze the cooling mechanism of human under exposure to EM radiation. Different thermal cooling mechanisms include blood flow, convective cooling, radiative cooling and their combinations are investigated at different ambient temperature. Numerical experiments show that convective cooling and radiative cooling rely heavily on the temperature difference between the body temperature and the ambient temperature. They are effective only when the ambient temperature is lower than the body temperature. Moreover, radiative cooling is more effective than convective cooling. Most importantly, the blood flow is very important for the thermoregulation of human body. It can increase the body temperature when the body temperature is lower than the core temperature and decrease the body temperature when the body temperature is higher than the core temperature. Especially when the body temperature is lower than the ambient temperature, the blood flow becomes the most effective cooling mechanism in human body.